# Spin flexoelectricity and chiral spin structures in magnetic films

A.P. Pyatakov[a]*, A.S. Sergeev[a], F.A. Mikailzade[b], and A.K. Zvezdin[c]

[a] M.V. Lomonosov Moscow State University, Leninskie gori, Moscow, 119991, Russia
[b] Department of Physics, Gebze Institute of Technology, Gebze, 41400 Kocaeli, Turkey
[c] A.M. Prokhorov General Physics Institute, Vavilova St., 38 Moscow, 119991, Russia

**Abstract**

In this short review a broad range of chiral phenomena observed in magnetic films (spin cycloid and skyrmion structures formation as well as chirality dependent domain wall motion) is considered under the perspective of spin flexoelectricity, i.e. the relation between bending of magnetization pattern and electric polarization. The similarity and the difference between the spin flexoelectricity and the newly emerged notion of spin flexomagnetism is discussed. The phenomenological arguments based on the geometrical idea of curvature-induced effects are supported by analysis of the microscopic mechanisms of spin flexoelectricity based on three-site ion indirect exchange and twisted RKKY interaction models. The electric-field-induced creation of the skyrmion-type structure is predicted.

## 1. Introduction

Rich diversity of chiral spin textures observed in thin films of magnets such as a long-wavelength modulation of magnetic order (spin cycloids) [1], homochiral domain walls [2],[3] and skyrmions [4–6] have inspired intensive research during the last few years. These topological objects demonstrate unusual properties promising for spintronics applications, such as a chirality-dependent current-driven domain wall motion [3], high sensitivity of skyrmions to spin-transfer torque exerted by a spin-polarized current [6], and an emergent feature of electric field induced magnetic domain wall displacement [7]. From the symmetry standpoints there is an analogy between the magnetic structures in thin films, the crystal subjected to flexural strain, the fan-shaped molecular structures in liquid crystals, and the spin cycloids in multiferroics (fig.1) [8]. Although the physical mechanisms stabilizing these topological textures are different in all these cases, the inversion symmetry breaking lifts the chiral degeneracy. In the following short review we will consider the physical origin of the chiral spin structures in magnetic films as well as various means to control them. The nascent fields of spin flexoelectricity [7] and flexomagnetism

* Corresponding author. Tel.: +7-495-939-4138; fax: +7-495-932-8820.
*E-mail address*: pyatakov@physics.msu.ru

[9] that deal with various chiral phenomena in magnetic media will be considered.

## 2. Spin flexolectricity and flexomagnetism

In thin films lack of space inversion symmetry leads to the interfacial internal electric field and the modification of their properties. The physical consequences are very similar to the ones observed on a curved surface or in a bent crystal [10] that is related to the symmetry equivalence of the flexural strain and polar distortion of the crystal lattice. Flexural deformation results in a distinct difference between the upper stretched layers and the bottom shrunk ones (fig. 1 a). This strain gradient singles out the polar direction in the crystal therefore its symmetry allows an electric polarization. The relation between flexural deformation and electricity appears in a flexoelectric effect in solid (fig.1 a) and liquid crystals (fig. 1 b) or in the *spin flexoelectricity* (fig.1 c), i.e. the electric polarization associated with the "bending" of magnetization distribution [7]. The fan-shaped spatially modulated structure shown in Fig. 1c is called the *spin cycloid*. In this context the spin cycloid observed in magnetic atom monolayers [1] was the result of the effective interfacial electric fields and can be attributed to the spin flexoelectric phenomena. The spatially modulated structures related to the spin flexoelectricity are not limited to the spin cycloid only but also include domain walls and skyrmions.

These spin textures are characterized by the direction of spin spatial rotation determined by the vector product of the adjacent spins $[\mathbf{s}_1 \times \mathbf{s}_2]$ that is frequently called "chirality" although the exact term is *vector chirality* (not to be confused with the chirality characterizing spin helicoid structure and magnetic vortices).

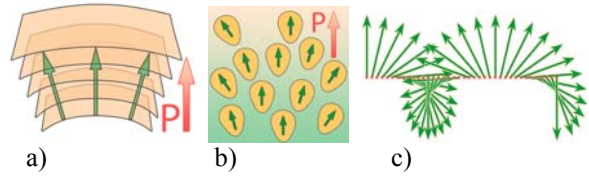

a)        b)        c)

Fig. 1. The symmetry analogy: a) flexoelectric effect in nonmagnetic solids b) molecular structures in liquid crystals c) spin cycloid structure in magnets.

Spin flexoelectricity is described by the contribution to the free energy in the form of Lifshitz-type invariant [11]:

$$F_L = \gamma \left( \mathbf{P} \cdot \{ \mathbf{m} \, div \, \mathbf{m} + \mathbf{m} \times curl \, \mathbf{m} \} \right), \quad (1)$$

where $\mathbf{m}$ is the unit vector of magnetic order parameter, $\gamma$ is flexomagnetoelectric constant, $\mathbf{P}$ is electric polarization or another polar vector (in the case of the thin film it is directed along the normal to the plane).

As will be shown below in Section 3 the Lifshitz-type invariant can be derived from the microscopic vector product of spins $[\mathbf{S}_1 \times \mathbf{S}_2]$. The switching of the vector chirality results in the sign change of spatial derivatives in (1) that leads to the reversal of electric polarization $\mathbf{P}$. The concept of spin flexoelectricity encompasses the broad range of magnetoelectric phenomena observed in magnetic ferroelectrics including manganites [12–14], tungstates [15], [16], and hexaferrites [17–20].

Spin flexoelectricity should not be confused with *flexomagnetism*, the term coined by R. Hertel [9] for a broad range of chiral phenomena observed in curved magnetic films. These geometry-induced effects do not necessarily involve electric field. For example, it was shown in [9] that the twist of flux-closure domain structures observed in curved soft-magnetic films can be reproduced by numerical micromagnetic simulation on curved surface without any special mechanisms or special types of chiral-dependent interactions. In these cases the origin of chiral phenomena is supposed to be the divergence of magnetization changing the magnetostatic energy.

In spite of the different mechanisms spin flexoelectricity and flexomagnetism have much in common: in both cases the spatial inversion breaking is the symmetry prerequisite of the chiral effects while the curvature appears either in the geometry of

## 3. Microscopic mechanisms of spin flexoelectricity

At the surfaces and interfaces the central symmetry of the lattice is broken that gives rise, in the presence of spin-orbit coupling, to an additional Dzyaloshinskii-Moriya interaction (DMI), i.e. the relativistic antisymmetric part of the Heisenberg Hamiltonian proportional to the cross product of the localized spins:

$$H_{DM} = \mathbf{D} \cdot [\mathbf{s}_1 \times \mathbf{s}_2], \qquad (2)$$

where $\mathbf{s}_1$, $\mathbf{s}_2$ are unit vectors of the magnetic moments of exchange coupled ions, $\mathbf{D}$ is the Dzyaloshinskii vector.

It is illuminating to compare the surface-related DMI with the DMI in weak ferromagnets (substances with noncollinear and not fully compensated magnetizations of antiferromagnetic sublattices). The same microscopic origin, i.e. the DMI (2), leads to the different macroscopic consequences:

(i)  a long range chiral spin structure
(ii) a homogeneous magnetic state with nonvanishing macroscopic magnetization.

The key factor here is the mutual direction of Dzyaloshinskii vector for adjacent pairs of ions. The constant value of $\mathbf{D}$ results in uniform spatial rotation of spin from ion to ion (fig.2 a) while alternating DMI restores the initial orientation of spin on every second ion (fig.2 b) thus resulting in the antiferromagnetic ordering with net magnetization due to the spin canting of magnetic sublattices.

In the conventional three-site indirect exchange model of DMI [21][22] these two cases have simple interpretation. In accordance to Keffer formula [23] the Dzyaloshinskii vector depends on the angle between the magnetic ion-ligand bonds:

$$\mathbf{D} = V_0 [\mathbf{r}_1 \times \mathbf{r}_2], \qquad (2)$$

where $V_0$ is microscopic constant, $\mathbf{r}_1$, $\mathbf{r}_2$ are the magnetic ions position vectors directed from ligand ion to the nearest magnetic ions (Fig. 2).

The uniform displacement of the ligands that is equivalent to the onset of polar direction in the crystal results in spin cycloid ordering (fig. 2 a) while the staggered ligand ion displacement leads to sign-alternating Dzyaloshinskii vector D and antiferromagnetic ordering with non-zero net magnetization (fig.2b).

Given the smallness of the displacement of the ligand ions $\mathbf{p}$ and spin canting $\delta\mathbf{s}=\mathbf{s}_1-\mathbf{s}_2$, the DMI term can be represented as

$$H_{DM} = V_0 ([\mathbf{p} \times \mathbf{a}] \cdot [\mathbf{s}_1 \times \delta\mathbf{s}]), \qquad (3)$$

where $\mathbf{a}$ is the lattice vector connecting the magnetic ions and $\mathbf{p}$ is the vector of polar displacement of the ligand. One can find by expanding the spin canting $\delta\mathbf{s}$ in Taylor series that the mixed type product of vectors (3) is the discrete form of the continual representation Eq.(1).

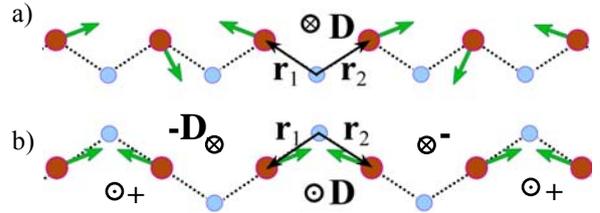

Figure 2 (color online). The three-site ion model for DMI. a) the unidirectional displacement of ligand ions shown as gray (blue) balls leads to the constant angle between the spins of magnetic ions shown as dark (brown) balls. b) the staggered displacements result in two antiferromagnetic sublatices with spin canting.

It should be noted that antisymmetric exchange is not limited by the three ion mechanism specific for magnetic dielectrics (oxides and fluorites). In magnetic metal films another scenario is possible based on the Ruderman-Kittel-Kasuya-Yosida (RKKY) exchange coupling in two dimensional electron gas (2DEG). RKKY interaction between localized spins is mediated by the conduction electrons whose spins in 2DEG systems precess due to the Rashba spin-orbit interaction. This precession leads to the spatial rotation of spin quantization axis as if we deal with the curved spin space. Note that although in this case it is supposed that Rashba interaction is caused by interfacial electric field there is another mechanism of the Rashba interaction [10] based on curvature-induced quantum effects for the case of bent magnetic nanostructures This shows that the analogy of 2DEG system with the curved surface is not limited to the specific DMI mechanisms.

In the presence of Rashba-field-induced spin space distortion the RKKY interaction in 2DEG is modified to the so-called *twisted RKKY interaction* [24].

$$H_{RKKY}^{twisted} = F(R)\mathbf{S}_1 \cdot \mathbf{S}_2(\theta), \quad (4)$$

where *F(R)* is the range function of RKKY coupling (the oscillating dependence of the exchange coupling on the distance), $\mathbf{S}_1$, $\mathbf{S}_2$ are the localized spins interacting via conduction electrons. In contrast to the case of conventional RKKY coupling the product of spins implies not only the scalar product of the spins but also the vector product:

$$\mathbf{S}_1 \cdot \mathbf{S}_2 = \cos\theta(\mathbf{S}_1\mathbf{S}_2) + \sin\theta[\mathbf{S}_1 \times \mathbf{S}_2]_y, \quad (5)$$

where $\theta = 2k_R(x_1 - x_2)$ (spins are supposed to be located in the positions $x_1$ and $x_2$ on the *x*-axis [24]), the $k_R$ is Rashba splitting of spin bands. For the case of 2D array of atoms the cross product of spins in the second term of (5) takes the form of the mixed product (3) provided that the polar vector **p** in (3) corresponds to the surface normal. The magnetic ion implantation in polar dielectrics [25] may provide new scenario for antisymmetric exchange and chirality-related spin flexoelectric interaction.

## 4. Chirality dependent domain wall motion

Although the first discovered homochiral surface spin texture was the spin cycloid observed in monolayers of Mn [1] the typical magnetic pattern for thin films is a domain structure with the spatial spin modulation in the domain walls separating homogeneously magnetized domains. Due to their mobility domain walls are sensitive to the various external influences like magnetic field, spin polarized current or even electric field. The chirality of the domain wall plays here an important role since it determines not only the velocity of the wall in magnetic field (as it happens in permalloy nanotubes due to their flexomagnetism [9]) but even the direction of the domain wall motion as will be shown in the examples below.

If magnetization rotates across the domain boundary in a cycloid-like way it is called a Néel-type domain wall (fig.3). Spin flexoelectricity according to (1) causes the electric polarization that creates the surface charges at the top and bottom of the domain wall providing a handle for the domain wall control with an electrically charged tip (fig.3a). The electrically induced magnetic domain wall motion in the rare earth iron garnet films [26,27] has been proved to be chirality dependent: the domain wall either attract to or repel from the tip in accordance with the direction of spin spatial rotation (Fig.3 b). The latter one is governed by an external magnetic field since the magnetization direction in the centre of the domain wall tends to align parallel to the field (fig. 3 a). The magnetic field reversal leads to the switching of the domain wall chirality and its electric polarity (fig.3 b) [27]. The adjacent domain walls move in opposite directions with respect to the tip (if one attracts to the electrode then the other repel from it, fig. 3b.).

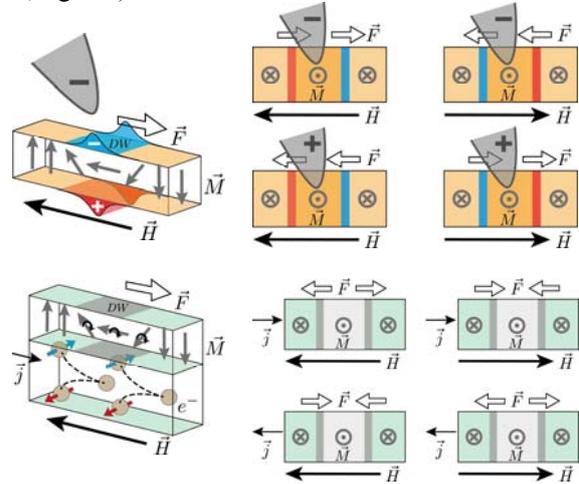

Figure 3. (Color online) Chirality dependent domain wall motion. The magnetic field *H* in all the cases is not the cause of the force *F* acting on the domain wall. *H* plays an auxiliary role switching the domain wall chirality. a) The electric polarization of the wall and surface charges due to the spin flexoelectricity b) the top view of the domain structure for both polarities of electric and magnetic field: force acting on the domain wall depends on the charge of the tip and the chirality of the wall (opposite chiralities shown with red/blue colors) [27]. c) Spin Hall effect in multilayers: the spin-Hall current injected from the bottom non-magnetic heavy metal layer produces spin-transfer torque in domain wall that can be described as a force acting on it d) the top view of the domain structure for various directions of current and magnetic field: as in the case of Fig.3b the neighboring domain walls have different chiralities [28].

Recently the chirality turned out to be a crucial feature in experiments on current-driven domain wall motion in heavy-metal/ferromagnet multilayers [3],[28]. In this case domain wall motion occurs due to spin-transfer torque acting on magnetic moments from spin-Hall current coming out of non-magnetic heavy metal layer (Pt, Ta, Ir) [29], as shown in Fig.3c. Direction of the force acting on the domain wall depends on the chirality of domain wall that results in motion of every second domain wall against an electron flow (fig.3c). This counterintuitive behaviour of the domain walls had been puzzling researchers until the chirality issue was considered [3].

Since the chirality determines the direction of the domain wall motion the way to control the spatial magnetization rotation should be found. To obtain a domain wall of certain chirality two schemes were proposed. The first approach is based on spin flexoelectric mechanisms considered in section 3. The central symmetry violation at the interface results in the homochiral domain wall structure [30], i.e. the direction of magnetization spatial rotation is identical at all domain boundaries and the spin vector rotates across the domain walls in a cycloid-like way. This type of structures were observed both in conducting and dielectric films of various thickness: in the double atomic layer of iron [2], in heavy-metal/ferromagnet/oxide nano-multilayers [3] and in epitaxially grown 10 μm-thick iron garnet films on gadolinium gallium garnet substrate [27]. In all these cases, the sense of magnetization vector rotation inside a domain wall is fixed by choice of material.

More flexible approach originally used in magnetoelectric study [27] and later in spin Hall experiments [28] that utilize external in-plane magnetic field aligning the magnetization in the center of the wall thus imposing certain direction of spatial magnetization rotation between two domains (fig.3 a,c). In this configuration neighboring domain walls have opposite chiralities that results in their opposite behavior under the influence of electric field and spin-Hall current (Fig. 3b, d).

### 5. Skyrmions and spin flexoelectricity

Skyrmion [4], i.e. the localized axisymmetric chiral spin texture (fig.4) stabilized by surface induced DMI [5] is now considered as a building block of future ultradense magnetic memory [6]. Skyrmion can be thought of as a domain wall coiled in a circle that suggests a possibility of all-electrical nucleation and stabilizing the skyrmion-type structure due to the spin flexoelectricity. We made corresponding numerical simulations with the use of simulated annealing technique of Ref.[4]. The two-dimensional model of magnetization distribution included inhomogeneous exchange $A=10^{-7}$ erg/cm, uniaxial out-of-plane anisotropy $K_u=10^3$ erg/cm$^3$ and spin flexoelectricity $\gamma=10^{-6}$ (erg/cm)$^{1/2}$ (the material parameters typical for iron garnet films [27]). The electric field was created by a capacitor with circular parallel plates with diameter of $7\Delta$, where $\Delta = (A/K_u)^{1/2}$ is the domain wall width parameter. Critical value of electric field strength required for skyrmion creation can be estimated as $10^6$ V/cm comparable to one used in experiments on iron garnet film [27].

Corresponding magnetization and polarization vectors distributions are shown in Figure 4. Note that polarization vectors have both vertical and radial components. First one corresponds to Neel-like structure of "coiled domain wall", and its sign is defined by the polarity of applied electric field. Second one arises due to in-plane curvature of the skyrmion.

This result is interesting in the context of recent reports on voltage-tuned DMI either through electric field induced electron density changes [31] or Rashba splitting modulation [32].

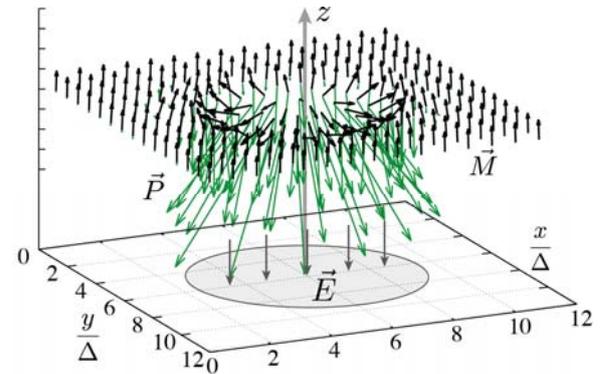

Figure 4. (Color online) Magnetization and electric polarization distribution in electrically stabilized skyrmion.

## 6. Conclusion

The geometrical idea of curvature which is present in all the considered cases (in the form of curved magnetic film, cycloidal magnetization distribution or twisted spin space) leads to nontrivial physical consequences and new phenomena like electrically stabilized spin cycloid structure and skyrmion, as well as electric field driven magnetic domain wall motion. In spite of the difference of physical mechanisms involved (electric field induced spin flexoelectricity, the magnetostatically governed flexomagnetism and spin torque driven domain wall dynamics) in all these cases the remarkable similarity in the physical implications is observed seen in the dependence of the domain wall motion on the chirality.

**Acknowledgements** The support from RFBR grant Support by Russian Foundation for Basic Research RFBR ## 13-02-12443 ofi_m2, 14-02-91374 ST_a, 14-29-08216 ofi_m is acknowledged.